\pdfoutput=1

\documentclass[aps,preprintnumbers,amsmath,amssymb,twocolumn, tightenlines,superscriptaddress]{revtex4}

\usepackage{graphicx}
\usepackage{epsfig}

\newcommand{\be}{\begin{eqnarray}}
\newcommand{\ee}{\end{eqnarray}}

 \newcommand{\gsim}{\mathrel{\hbox{\rlap{\lower.55ex \hbox {$\sim$}}
                   \kern-.3em \raise.4ex \hbox{$>$}}}}
\newcommand{\lsim}{\mathrel{\hbox{\rlap{\lower.55ex \hbox {$\sim$}}
                   \kern-.3em \raise.4ex \hbox{$<$}}}}

\newcommand{\ba}{\begin{eqnarray}}
\newcommand{\ea}{\end{eqnarray}}

\begin{document}


\title{The Sounds of the QCD Phase Transition}
\author{Edward Shuryak and Pilar Staig}
\address{Department of Physics and Astronomy, 
Stony Brook University, NY 11794, USA.}

\begin{abstract}
Hydrodynamic description of a fireball produced in high energy heavy ion collisions has been recently supplemented by a very successful study of acoustic perturbation  created by the initial state perturbations. We discuss sound  produced at later stages of the collision, as the temperature drops below critical, $T<T_c$, and originated from the Rayleigh-type collapse of the QGP clusters.  In certain  analytic approximation we study distorted ``sound spheres" and calculate modifications of the particle spectra and two-particle correlators induced by them. Unlike for initial state perturbations studied previously, we propose to look for those late-time sounds using rapidity correlations, rather than the azimuthal angles of the particles. We then summarize known data on rapidity correlations from RHIC and LHC, suggesting that the widening of those can be the first signature of the late-time sounds.
\end{abstract}

\maketitle
    \section{Introduction}
Production of Quark-Gluon Plasma (QGP) has been a major goal of the heavy-ion collision program at RHIC and LHC. Relativistic hydrodynamics describes quite accurately the radial and elliptic flows seen in RHIC data
\cite{Teaney:2000cw,Teaney:2001av,Hirano:2004ta,Nonaka:2006yn}.

The observed explosion has certain similarities with the cosmological Big Bang, and is often called ``The Little Bang". This analogy extends to the existence in both cases of small deviations from the smooth average behavior caused by quantum fluctuations at the early stages of the process. Resulting   event-by-event fluctuations of the elliptic flow  \cite{Mrowczynski:2002bw}, as well as   the third \cite{Alver:2010gr} and higher harmonics are also well described by hydrodynamics, as shown in multiple papers  (including those by the current authors  \cite{Staig:2010pn,Staig:2011wj}). One important conclusion from all those works is that the mode damping is basically acoustic in nature, and consistent with the same value of the viscosity-to-entropy ratio $\eta/s=(1.5-2)/4\pi$
see e.g.  a recent phenomenological summary   \cite{Lacey:2013is}. 
The  quantum fluctuations of the colliding nuclei are,  however, neither the only nor even the most interesting source of fluctuations. The necessary existence of fluctuations $during$ the hydrodynamical expansion follows from dissipation-fluctuation theorem and its theoretical grounds have been recently studied by Kapusta, Muller and Stephanov \cite{Kapusta:2011gt}. As the system expands and its temperature passes through the phase transition region $T\approx T_c$, from QGP to the hadronic phase, one may think of enhanced critical fluctuations
 \cite{Shuryak:1997yj} . Those  are expected to be enhanced 
near the second-order critical point  \cite{Stephanov:1998dy}: this idea had
motivated the so called  downward energy scan program at RHIC, not yet completed.
  
In this paper we propose a different strategy in a search for the critical  event-by-event fluctuations: using the $sound$  emitted by them.  The very strong interaction in the system, leading to a rapid relaxation, from an enemy becomes an ally. In a near-ideal fluid  the sounds are the only long-lived propagating mode. The underlying assumption is that the ``acoustic" properties of the matter are there not only during the QGP era, but are also maintained for the time period between the critical region ($T\approx T_c\approx 170\, MeV$) and the final (kinetic) freeze-out    ($T\approx  100\, MeV$) . 

The sound generation by critical fluctuations while crossing the phase transitions is a well known phenomenon in physics: e.g.  familiar   ``singing" of a near-boiling tea pot. While the QCD phase transition is not strictly a first order transition, but a smooth cross-over, it is still close to it. At certain expansion rate of the fireball,  formation of an inhomogeneous intermediate state in the near-$T_c$ region is quite probable. By its end,  certain QGP clusters should remain. (Even pp collisions result in significant clustering of secondaries, as two-particle correlations in rapidity indicate.) A new idea of this paper is that, instead of slowly evaporating, the QGP clusters  should undergo Rayleigh-type collapse, transferring (part of) their energy/entropy  into the outgoing shocks/sounds.

These ``mini-Bangs", as we will call them, are the source of the sound spheres, distorted by flow.   In order to separate them from sounds caused by the initial state perturbations, one may use their early-time origins and rapidity independence.  The late-time  ``mini-Bangs" have also sound waves propagating in longitudinal(rapidity) direction. As we will show in this paper, some of them should produce correlations rather different from the usual Gaussian-like correlations, coming from isotropic resonance/cluster decays.

As the trigger $p_t$ grows  the contribution of the jet fragmentation also grows, and  beyond say 10 GeV it becomes dominant. Whatever the model of jet quenching, it is clear that some fraction of the energy goes into the medium and thus jets must also  induce a sound wave \cite{CasalderreySolana:2004qm}.  From a hydrodynamical point of view, these sounds are similar to those from the ``mini-bangs", and differ only by the fact that jet quenching deposits energy along the light-like trajectory rather than at a particular space-time point. We will not discuss sounds from jets in this paper, as we do so elsewhere \cite{sounds_from_jets}.

The structure of the paper is as follows. Section \ref{sec_drops} describes cluster implosion in the passing through the phase transition. It starts with a brief review of the  Rayleigh collapse phenomenon, known for a long time and studied extensively in sonoluminescence experiments. Our studies focus on the shock/sound generation and the role of viscosity. 

In section \ref{sec_sounds} we study sound propagation, on top of the expanding fireball. We are fortunate to be able to do it near-analytically, using a small perturbation on top of the so called Gubser flow solution. We end up calculating the shapes of the two-particle correlations those sounds produce. In the next section \ref{sec_pheno} we compare the results with the ALICE (LHC) data, which indeed show the double-hump correlations we propose to identify with the ``mini-Bangs" .

\section{Cluster collapse, shock/sound formation}
\label{sec_drops}
\subsection{The Rayleigh collapse}

This subsection  contains well known material worked out by people working on sonoluminescence, for a review see e. g. \cite{Brenner:2002zz}: it is given for self-consistency of the paper, introduction of notations etc.

We start by reminding the derivation of the Rayleigh equation for the bubble radius, coming from the Euler hydro equations
\ba 
\rho [\partial_t \vec u+ (\vec u \vec \nabla) \vec u] & = & -\vec \nabla p \, , \nonumber \\
\partial_t \rho+\vec \nabla(\rho \vec u) & = & 0 \, .
\ea
The  standard steps are the assumption of spherical symmetry of the flow, and the introduction of the flow potential 
\ba  \vec u=   \vec\nabla \phi(r,t)\, . \ea
Then, stripping off the gradient, one finds  that the first Euler equation becomes
\ba \rho \partial_t \phi+ \frac{1}{2}(\partial_r \phi)^2=-p \, . \label{eqn_E2}\ea
Using $dp/d\rho=c^2, dh=dp/\rho$  where $h$ is the enhtalpy, and $c$ is the sound velocity (the speed of light in our units is 1), one obtains a single equation for flow potential
\ba 
 \vec\nabla^2 \phi- {1\over c^2} \partial_t^2\phi=  {u\over c^2}(  \partial_t u- \partial_r h) \, .
\ea

Now comes the crucial step: if all flows are slow compared to $c$,  the Laplacian term is the dominant one.
 It then provides a simple Coulomb-like solution to the potential \be \phi\sim f_1(t){1 \over r} +f_2(t)\, , \ee 
as a function of $r$. The two  time dependent functions should be matched to the boundary conditions of the problem. One of them is at the bubble wall located  at some $R(t)$:  the condition matches the flow velocity with the wall speed
\ba
u_r=\partial_r \phi=\dot{R} \, , \ea
where the dot denotes the time derivative. It fixed one of the function in  a solution 
\ba 
\phi=-{ \dot{R}  R^2 \over r} + f_2(t)\, ,
\ea
and putting it back into Euler equation in the form (\ref{eqn_E2}) one finds, taking at $r=R$, the ordinary differential equation for $R(t)$
\ba 
 \ddot{R} R + (2-1/2) \dot{R}^2 ={p(r\rightarrow\infty,t) \over \rho}\, ,
\ea
where the (1/2) comes from the second term of  (\ref{eqn_E2}) and the r.h.s. is the effective pressure far from the bubble. 

If the r.h.s. is positive, the system is stable, but as it crosses into the negative a collapse takes place.
What was discovered by Lord Rayleigh is that even if the r.h.s. is
put to zero, the equation admits a simple analytic solution (known as  the original Rayleigh collapse solution)
\be R(t) \sim (t_*-t)^{2/5} \, .\ee
While the time-dependent singularity has a positive power, it is less than one, and thus produces
an {\em infinite velocity} \be \dot{R}\sim  (t_*-t)^{-3/5} \, ,\ee at $t=t_*$. 
Needless to say, large velocity is incompatible with the approximation of small $u<<c$ made above: therefore  the near-collapse stage should be treated separately and more accurately (see below).

A comprehensive  review \cite{Brenner:2002zz} on sonoluminiscence includes both the theoretical and the phenomenological discussion of the shock waves produced by the collapsing air bubbles in water, under the influence of small-amplitude sounds driving an effective pressure to negative at each sound cycle. The reader interested in details can find it in this review: let us only mention that
the observed shocks from collapsing bubbles have  velocities  of about 4 km/c,  few times the speed of sound in water $c=1.4\,  km/s$,  suggesting the pressure in the collapse reaching a range as high as 40-60 kbar.
Those values also imply a reduction of the bubble's volume by a huge factor $\sim 10^6$. Emission of light, indicating very high temperatures $T\sim 1 \, eV \gg T_{r\rightarrow\infty}$, gave the name of $sonoluminiscence$ to the whole phenomenon. Last comment is that in these experiments one found  a rather high efficiency $\sim O(1/2)$ of the energy transferred into the shocks/sounds.

\begin{figure}[!t]
 \center{ \hskip 0in\includegraphics[width=7cm]{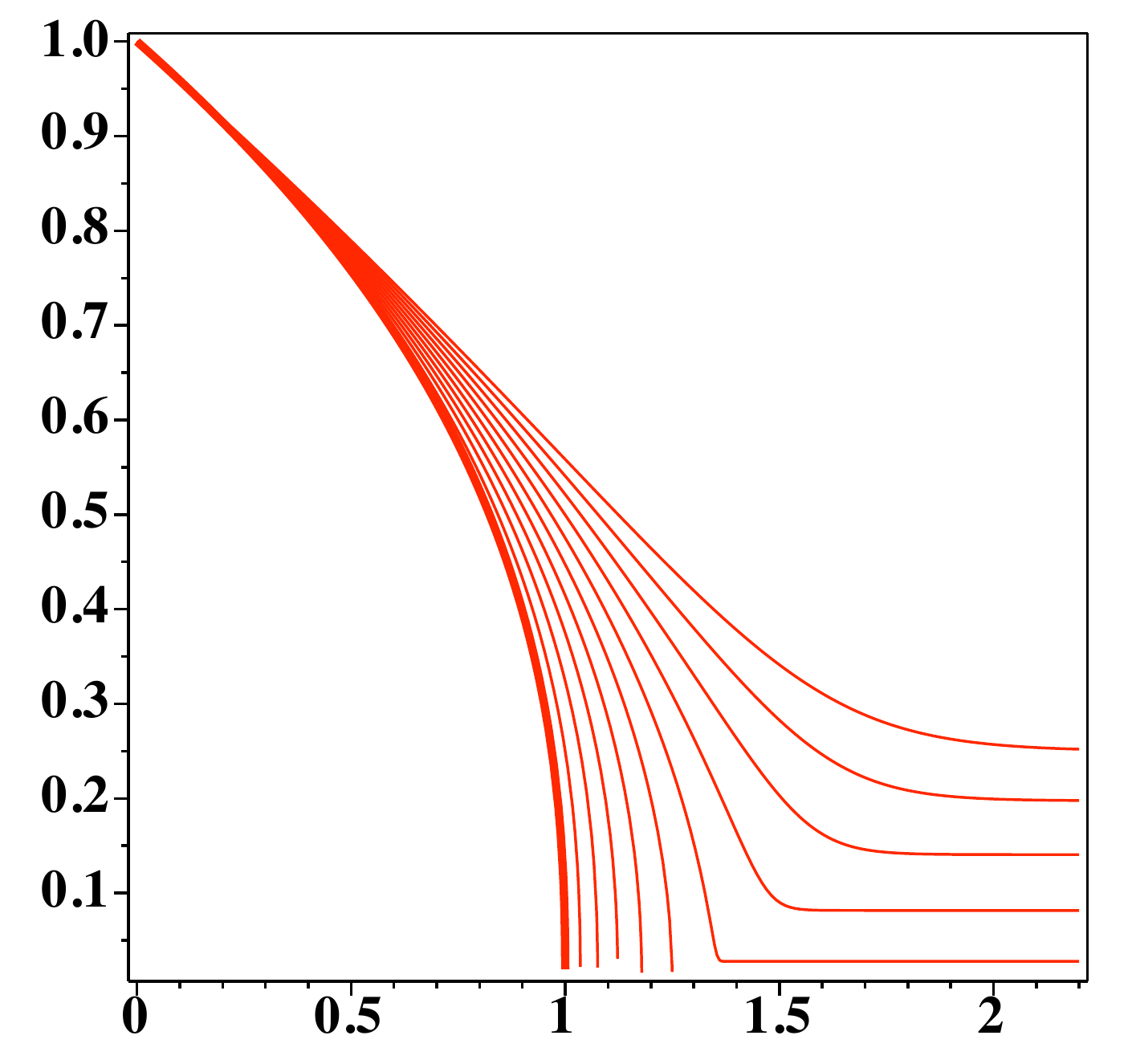}}
 \caption{
 The time evolution of the drop radius $R(t)$, for the values of $\eta/\rho=0.01..0.1$ with a 0.01 step.
 \label{fig_collapse_eta} }
 \end{figure}

\subsection{The collapse with the viscosity and  sound radiation}
The r.h.s. of the equation for the $R(t)$ can include a number of extra terms. The most obvious of them is the bulk pressure, which drives the collapse. The next is the surface tension, preventing collapse of too small bubbles because its role grows as $1/R$ at small $R$. Ignoring those terms for now, we focus on 
 the dissipative effect of the flow due to viscosity. Standard  Navier-Stokes term in the r.h.s. is
\ba
\ddot{R} R + \frac{3}{2} \dot{R}^2 =-{4 \eta \dot{R} \over \rho R}\, .
\ea
Solving this equation with variable value of the viscosity we found its critical magnitude capable to turn the catastrophic Rayleigh collapse into a ``soft landing". In Fig.\ref{fig_collapse_eta} we  show a set of  solutions with increasing values of $\eta*T/\rho$, showing how  the collapse can be stopped by viscosity. The value of the ratio $\eta*T/\rho>0.6$ is needed for this to happen. 
For smaller values it goes into the Rayleigh singularity, which simply stops our numerical solver (we use default one on Maple 16). 

The second effect we  study  is the sound radiation. For a spherical source with a time-dependent volume $V(t)=(4\pi/3) R(t)^3$ the outgoing wave solution at large distances is (see hydrodynamics textbooks such as \cite{LL_hydro}) 
\ba   \phi= -{\dot{V}(t-r/c) \over 4\pi r} \, , \ea
corresponding to the flow velocity of radiated sound 
\ba v_r=\dot{R}={\ddot{V} \over 4\pi r c} \, ,\ea 
resulting in the intensity of the sound radiation 
\ba 
I= {\rho \over 4\pi c} |\ddot{V}|^2 \, ,
\ea
at large distances. In Fig.\ref{fig_collapse_eta} we plot the time evolution of the volume acceleration squared (to which sound radiation intensity
is proportional) for five trajectories, generated by smooth {\em viscosity-induced} end of the collapse. What one can see from those figures is that the sound radiation has a sharp peak at certain moment, which becomes much more pronounced  as the viscosity is reduced toward its critical value mentioned above.  This peak in the sound emission represents  the ``mini-bang" we are discussing in this paper.

\begin{figure}[!t]
 \center{ \hskip 0in\includegraphics[width=7cm]{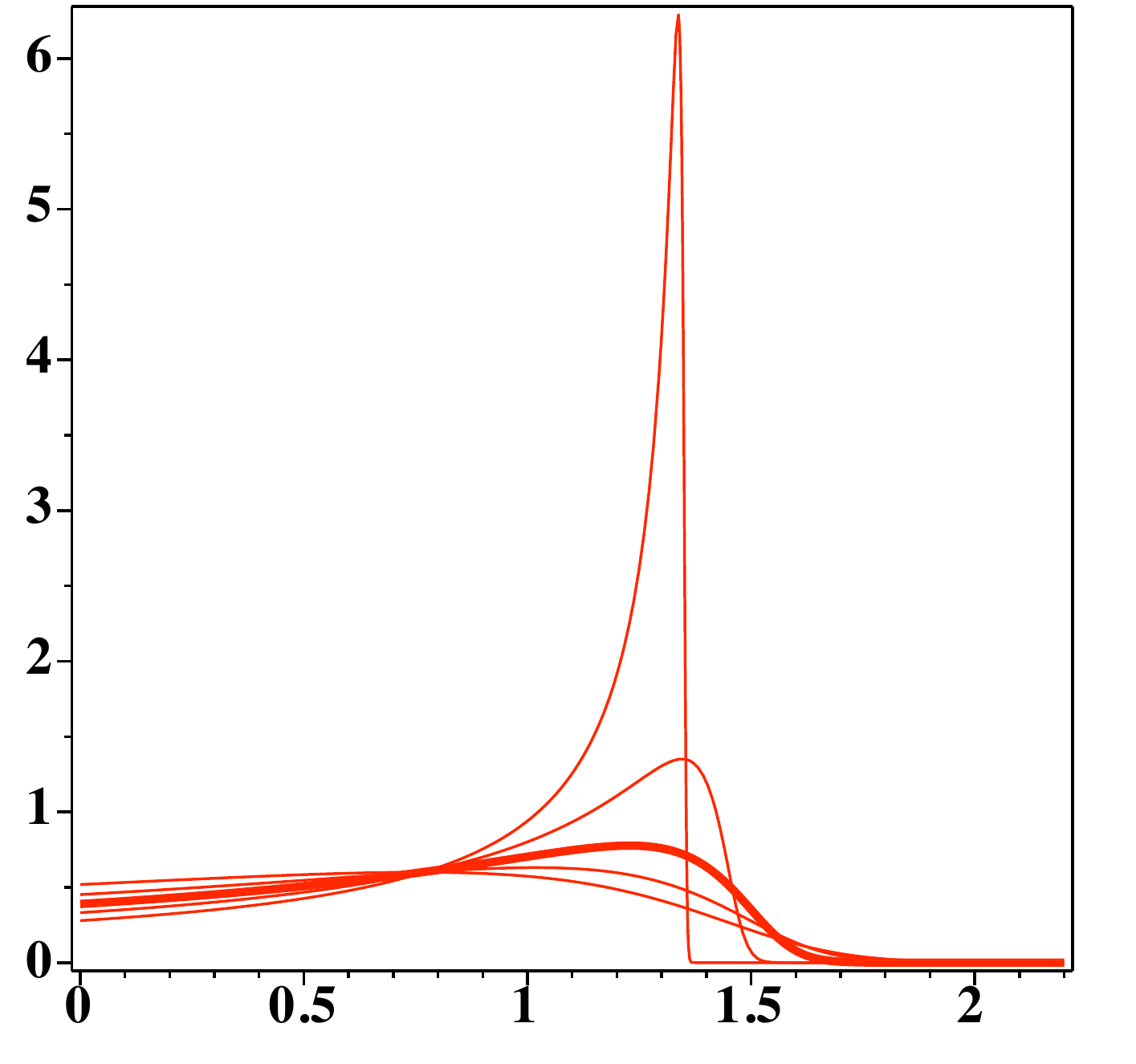}}
 \caption{
 The time evolution of the quantity $|\ddot{V}(t)|^2$, entering the sound radiation intensity, for the values of $\eta/\rho=0.06,0.07,0.08,0.09,0.1$.
 \label{fig_collapse_eta} }
 \end{figure}

It is methodically interesting (see refs in \cite{Brenner:2002zz}) to derive the ``self-force" induced by the sound radiation directly, which is analogous to the Abraham-Lorentz reaction-to-radiation force in electrodynamics. Including in $\phi$ the outgoing sound one determines the second function of the time
\ba 
\phi= \phi_\infty(t)-{1\over r} F(t-r/c)\approx  \phi_\infty(t)-{1\over r} F(t)+ {1\over c} \dot{F}\, ,
\ea    
where, as before $F(t)=\dot{R} R^2$, one finds a contribution to the r.h.s. of the main equation to include the $third$ order derivatives of the radius 
\ba
\rho (\ddot{R} R + \frac{3}{2} \dot{R}^2)=...+ {\rho \over c} {d^2 \over dt^2 }( {dR\over dt} R^2)\, ,
\ea
similar to the familiar Abraham-Lorentz one. (The only difference really is that the dipole radiation in electrodynamics is substituted by spherical monopole radiation of sound.) As it is the case with other self-force applications, one needs more initial conditions. Also having small terms with higher derivative prone to spurious a-causal solutions, so this equation is to be treated with care.  Yet using the Rayleigh collapse solutions with viscosity we already have, one can calculate this term and see that it is indeed very singular. Perhaps one needs higher derivatives and nonlinear equations to get more accurate solution near the singularity: yet the main answer is clear: the energy of the collapsing  

\section{Sound propagation  on top of expanding fireball}
\label{sec_sounds}
\subsection{General considerations}
It is by now well established that the 4-dimensional region of space-time in which hydrodynamical description is (approximately) valid is surrounded by the 3-dimensional surface, consisting of the ``initial" and ``final" parts, in which  the signs of the matter flow through it are in and out, respectively. The observed secondaries come from the latter part, and their spectra are commonly calculated by the  Cooper-Frye formula \cite{Cooper:1974mv}
\begin{eqnarray}
\frac{dN}{dy p_T dp_T d\phi} & = & \int_{\Sigma}
f(p^{\mu}u_{\mu})p^{\mu}d^3\Sigma_{\mu}\, , \label{eqn_Cooper_Fry}
\end{eqnarray}
where $p^{\mu}$ is the 4-momentum of the particle, $u_{\mu}$ the 4-velocity of the fluid, $d\Sigma_{\mu}$ is the vector normal to the freeze-out surface and the function $f$ is the distribution function of particles that we approximate by a Boltzmann distribution.

\begin{figure}[!t]
\includegraphics[width=7cm]{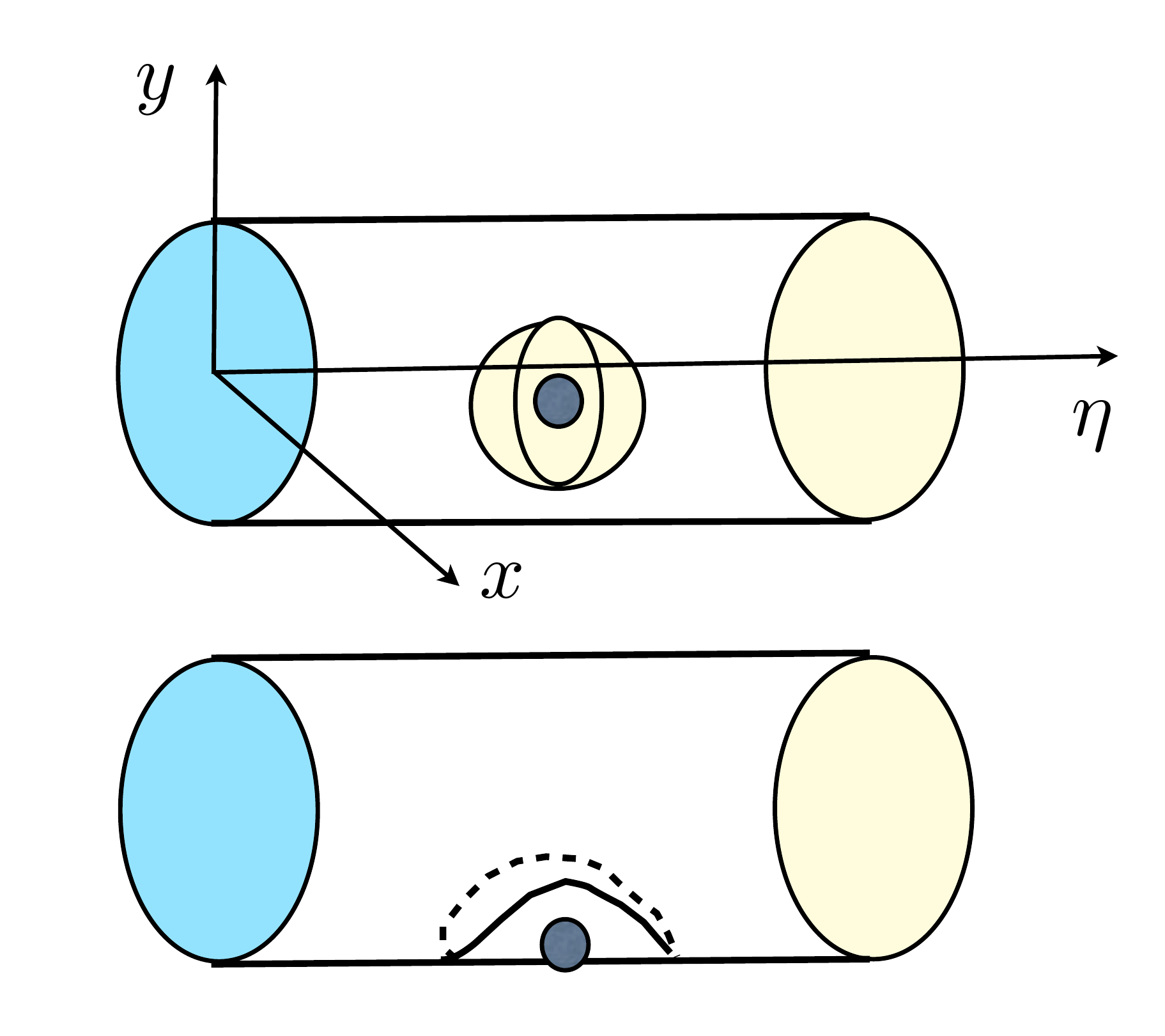}
 \caption{The schematic representation of the freeze-out cylinder and a sound perturbations, for ``internal"
(upper picture) and ``peripheral" (lower picture) sources. In the latter case we show the intersection of the sound sphere and the  freeze-out cylinder, called
in the text the ``sound line". }
 \label{fig_cyl} 
 \end{figure}

For secondaries with the $p_t$ of interest -- say 1.2 to 2.4 $GeV$,  which are well described by hydrodynamics -- such $p_t$ exceeds the freeze out temperature $T_f\approx 120\, MeV$ by a large factor ranging from 10 to 20.  If those particles were produced by the pure tail of the thermodynamic Boltzmann factor, its probability would be truly negligible.  But  the hydrodynamical expansion makes  a huge difference: in the moving fluid  the exponent is not the energy in the lab frame but in the frame $comoving$ with the fluid, $p^\mu u_\mu$, which is smaller than the momentum itself by the so called ``blue shift factor" 
\be  {p^\mu u_\mu \over T_f}\approx {p_0 \over T_f} \sqrt{{1-v_\perp \over 1+v_\perp}} \, . \label{eqn_param}  \ee
It depends on the local transverse flow velocity $v_\perp$ which varies over the surface $\Sigma$, with a maximum   near the edge. The transverse flow velocity reaches at LHC $v_\perp \sim 0.8$, for which this factor is $\sim 1/3$, reducing the quantity in Boltzmann's exponent to only  $\sim 3..7$. It is much smaller than $p_t/T_f$, but still can be considered a large parameter. This blue shift narrows the contribution from  the surface integral to the particle spectra to relatively small vicinity of the point $r= r_*$ where $r_*$ is the location of the {\em maximal transverse flow}. (We will specify it in the next section using a particular analytic example.)  Furthermore, assuming for simplicity zero impact parameter (central collisions) and rapidity independence of the system, we conclude that at such $p_t$  the observed particles come  from the 
``freezeout cylinder", with $r=r^*$, depicted in Fig.\ref{fig_cyl}. Large transverse flow strongly enhances the contribution of this cylinder, basically projected it onto the detector. 

\begin{figure}[!t]
\includegraphics[width=7cm]{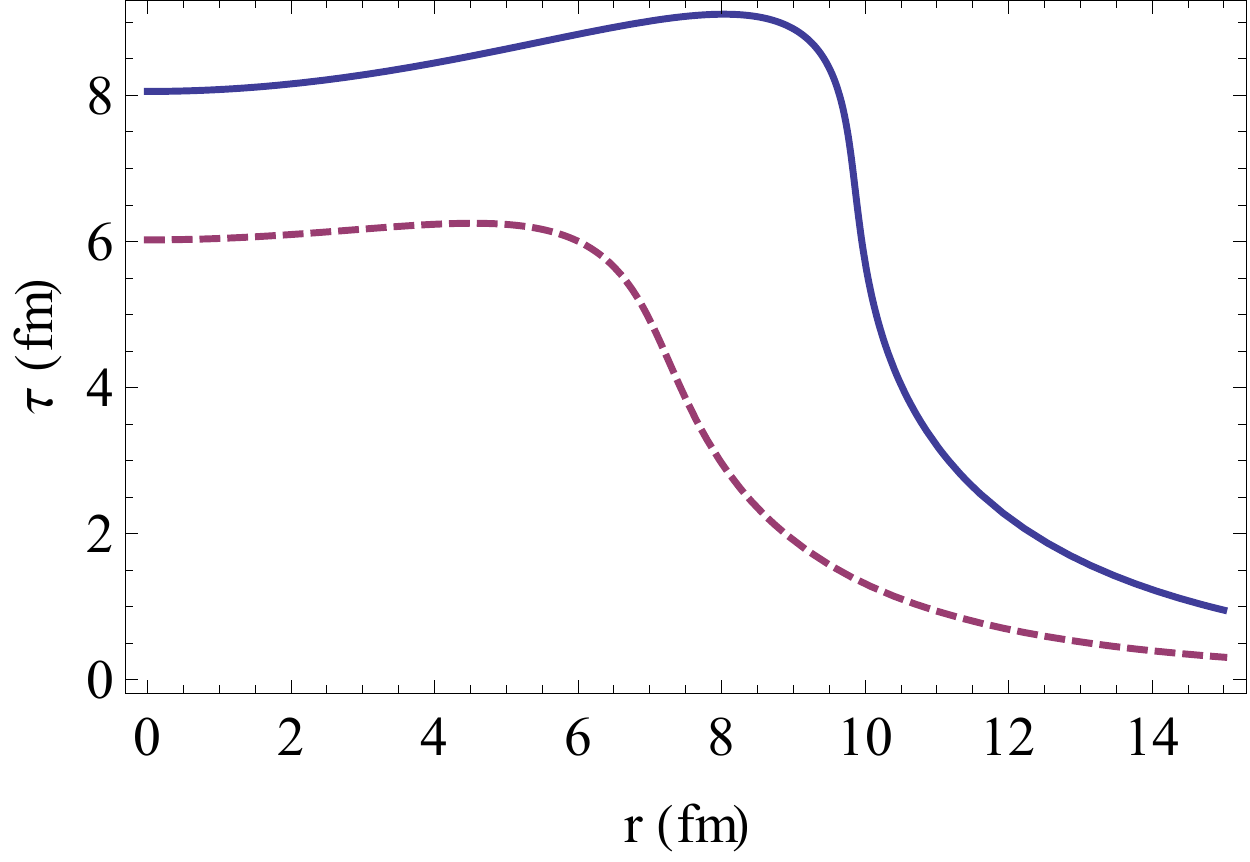}
 \caption{Isothermal surfaces for $T_c=175$ MeV (blue solid line) and $T_{fo}=120$ MeV (magenta dashed line). The sound emission from near-$T_c$ phenomena is expected at the former surface, and its propagation ends on the latter one. }
 \label{tauc_taufo} 
 \end{figure}

Now let us consider small-amplitude  sound perturbations, propagating on top of the background flow. They form certain distorted sound sphere around the origination point. From the discussion above it is clear that such perturbations fall into two classes: (a) the ``internal" ones, such that their sound sphere never reaches the flow maximum on the freezeout surface, and (b) the ``peripheral" ones, for which the sound sphere and the 
freeze-out cylinder cross, see Fig. \ref{fig_cyl}. From the previous discussion the latter perturbations should be dominant over the former, as they benefit maximally from the blue-shift effect. Thus we come to conclusion that  clusters located not too far from $r=r^*$ cylinder are the only ones which can be  observed. 
The distance from it is given by the distance the sound can travel between its emission point and the final freeze out moment.  Thus the corresponding perturbation should be located approximately at a ``sound sphere" of the radius $R_s=c_s (\tau_f-\tau_{emission})$, distorted by the flow, shown schematically in   Fig. \ref{fig_cyl}(b).  As one can see from Fig.\ref{tauc_taufo}, the time difference between the two surfaces is about $2\,  fm/c$ for $r_{emission}<6 \, fm$, but grows to 6 fm/c at $r_{emission}\approx 8 \, fm$. In the former case the  radius of the sound sphere is about $R_s\sim 1\, fm$ in absolute distance: compared to the size of the   fireball one finds the expected angle $\Delta \phi\approx  R_s/r^*\sim 1/7 $: too small, well inside the peak of comoving particles of jets and mini jets. However if the sound source is at the ``outer wall", with $r\approx 8 \, fm$, the time and corresponding angles $\Delta\phi,\Delta \eta$ are larger and may become observable. Repeating the same argument as above, we expect that the observable effect is basically a projection of the  intersection of the sound sphere and the freeze-out cylinder, where both the perturbation and blue-shift are maximal.
  
Summarizing this section:  the sounds emitted  close to the fireball surface are most likely to be detected. The best  observable correlations induced by the sound come from the  intersection of the sound sphere and the freeze-out cylinder. In principle, one should find effect both in azimuthal and rapidity directions.
 
\subsection{Perturbations of the Gubser's flow}
When the radius of a cluster rapidly decreases, it produces spherical sound waves that expand and propagate through the medium.  This effect is similar to what happens when a Gaussian perturbation is placed in the medium: it too generates divergent sound waves.  We will now look at the effect that the presence of such waves  has on  the final particle distribution. 

Hadronic matter is different from QGP in the speed of sound: $c^2\approx 1/5$ rather than $1/3$. We will however ignore this difference and use the SO(3)-invariant flow developed by Gubser and Yarom \cite{Gubser:2010ze,Gubser:2010ui} as a background. We only want in this work to have a qualitative description of the propagating sounds, and Gubser-Yarom framework provides very nice analytical tools to do so. Furthermore, a propagation of perturbation induced by a Gaussian source we had already studied in \cite{Staig:2011wj}. Two new elements are: (i) the perturbation is not defined at initial time, but at some ``hadronization" surface; (ii) instead of propagation in 3 dimensions as before, we now consider all 4 dimensions,  including spatial rapidity $\eta$.   In this framework it is useful to work in the $(\rho,\theta,\phi,\eta)$ coordinates, related to transverse radius $r$ and proper time $\tau$ by
\begin{eqnarray}
\sinh{\rho} & = & -\frac{1-q^2\tau^2+q^2r^2}{2q\tau}\, ,\\
\tanh{\theta} & = & \frac{2qr}{1+q^2\tau^2-q^2r^2}\, ,
\end{eqnarray}
where $q$ is the dimensional parameter giving the size of the fireball. 
These coordinates are comoving coordinates in the background flow, in which hydrodynamics of perturbations
allow for separation of all 4 coordinates. Moreover, 
the azimuthal angle $\phi$ and $\theta$ are combined together into those on 2-sphere, so the corresponding set of functions are just standard spherical ones $Y_{lm}(\theta,\phi)$. 

The temperature and velocity are given by the general expressions
\begin{eqnarray}
T & = & \frac{T_0}{\tau \left( \cosh{\rho}\right)^{2/3}}\left( 1 + \right. \nonumber\\ 
&{} & \left. \sum c_{k l m}\delta_{kl}(\rho)Y_{l m}(\theta,\phi) e^{i k \eta}\right) \, ,\\
u_{\tau} & = & u_{\tau,Back} + \nonumber\\
& {} & \tau \frac{\partial \theta}{\partial \tau}\sum c_{k l m}v_{kl}(\rho)\partial_{\theta}Y_{l m}(\theta,\phi) e^{i k \eta}\, ,\\
u_{r} &  = & u_{r,Back} +  \nonumber\\
&{}& \tau \frac{\partial \theta}{\partial r}\sum c_{k l m}v_{kl}(\rho)\partial_{\theta}Y_{l m}(\theta,\phi) e^{i k \eta} \, ,\\
u_{\phi} & = &\tau \sum c_{k l m}v_{kl}(\rho)\partial_{\phi}Y_{l m}(\theta,\phi) e^{i k \eta}\, ,\\
u_{\eta} & = & \tau \sum c_{k l m}v_{kl}^{\eta}(\rho) Y_{l m}(\theta,\phi) e^{i k \eta}\, . 
\end{eqnarray}

\begin{figure}[!t]
\includegraphics[width=8cm]{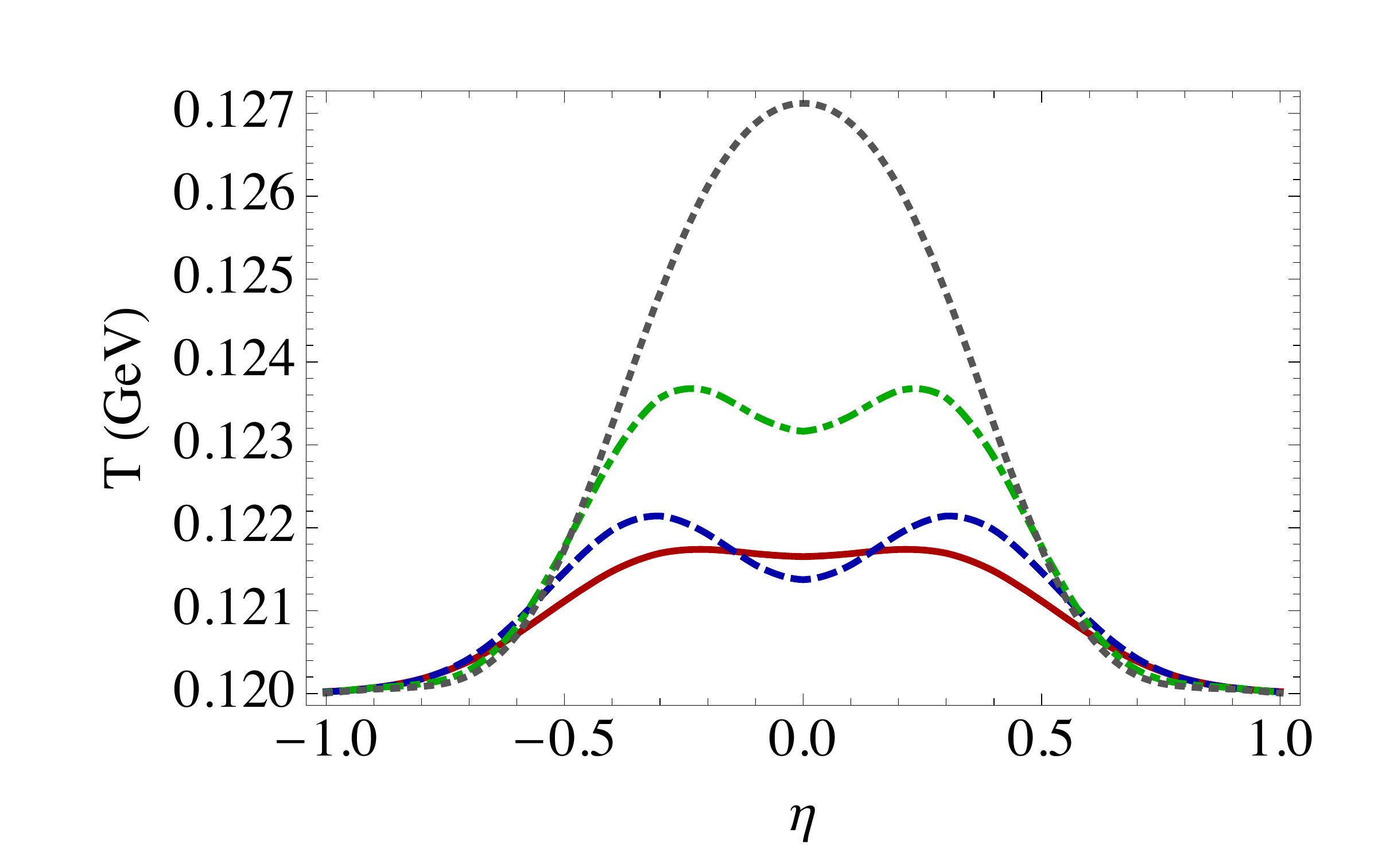}
 \caption{ Temperature at the (zeroth order) freeze-out as a function of spatial rapidity $\eta$, for transverse plane coordinates $ x=6, 7, 8, 8.3 fm,y=0 \, fm,$  (red, blue, green, gray, respectively).}
 \label{fig_deltaT} 
\end{figure}

The background flow ($u_{\tau,Back}$ and $u_{r,Back}$) is described by the (axially symmetric) flow  proposed
 in \cite{Gubser:2010ze,Gubser:2010ui}, the constants $c_{k l m}$ were calculated by imposing that the pertrubation starts as a Gaussian in $\theta,\, \phi$, and $\eta$, and the $\rho$-dependent functions $\delta(\rho), \, v(\rho)$ and $v_{\eta}(\rho)$ were computed from the system of coupled differential equations eq.(108,109)  in \cite{Gubser:2010ui}.
We placed the perturbation near the edge of the expanding matter at the time when the medium reaches the critical temperature, and let it evolve until the system reaches freeze-out. A sample of results is shown in Fig.\ref{fig_deltaT},
corresponding to different cuts through the ``sound sphere".
One can see that at appropriate positions the double-peak structure in longitudinal coordinate -- represented by a spatial rapidity $\eta$  --  emerges, substituted by a single peak centered at the cluster rapidity (gray dotted line) when looking at the very edge of the fireball.

\subsection{Particle spectra and correlations}
As it is by now well known, different species of secondaries are affected by collective flow differently. One consideration, stemming from previous studies of elliptic and  higher harmonics of the flow, is that the largest effects are observed at $p_\perp \sim 3\, GeV$, where those effects are maximal. 
The second general consideration is that the heavier the particle used, the smaller its thermal motion inside the cell at freeze out, and thus the more visible collective velocities become. In particular, the  baryons/antibaryons have thermal velocities  $\sim \sqrt{T_f/M}\sim 1/3$, significantly less than
pions. Note also, that using protons instead of pions does not result in significant loss of statistics, as at transverse momenta range under consideration their spectra are comparable.   

To calculate the final particle distributions we used an approximate isothermal freeze-out prescription, taking as freeze-out surface the surface obtained by setting $T_{Back}(\tau,r) = T_{fo}$ (solid curve in Fig. \ref{tauc_taufo}).  
We then computed the Cooper-Frye integrals (\ref{eqn_Cooper_Fry}) to get the particle distributions.  

As explained in detail in \cite{us_pt_paper}, the space-like part of the freeze out surface for Gubser flow deviates significantly
from the one obtained in more realistic numerical hydrodynamical simulations. 
The relatively long power-like tail of the matter distribution at large distances do not correspond to exponential cutoff
of the density at the edge of the nuclei.
Fortunately, the realistic freeze out surface normal is nearly orthogonal to momenta and thus it contributes
only few percents to the final spectra. Thus we adopted a simple  practical approach: we simply ignore it and
include only the contribution of the time-like part. 

\begin{figure}[!t]
\includegraphics[width=7cm]{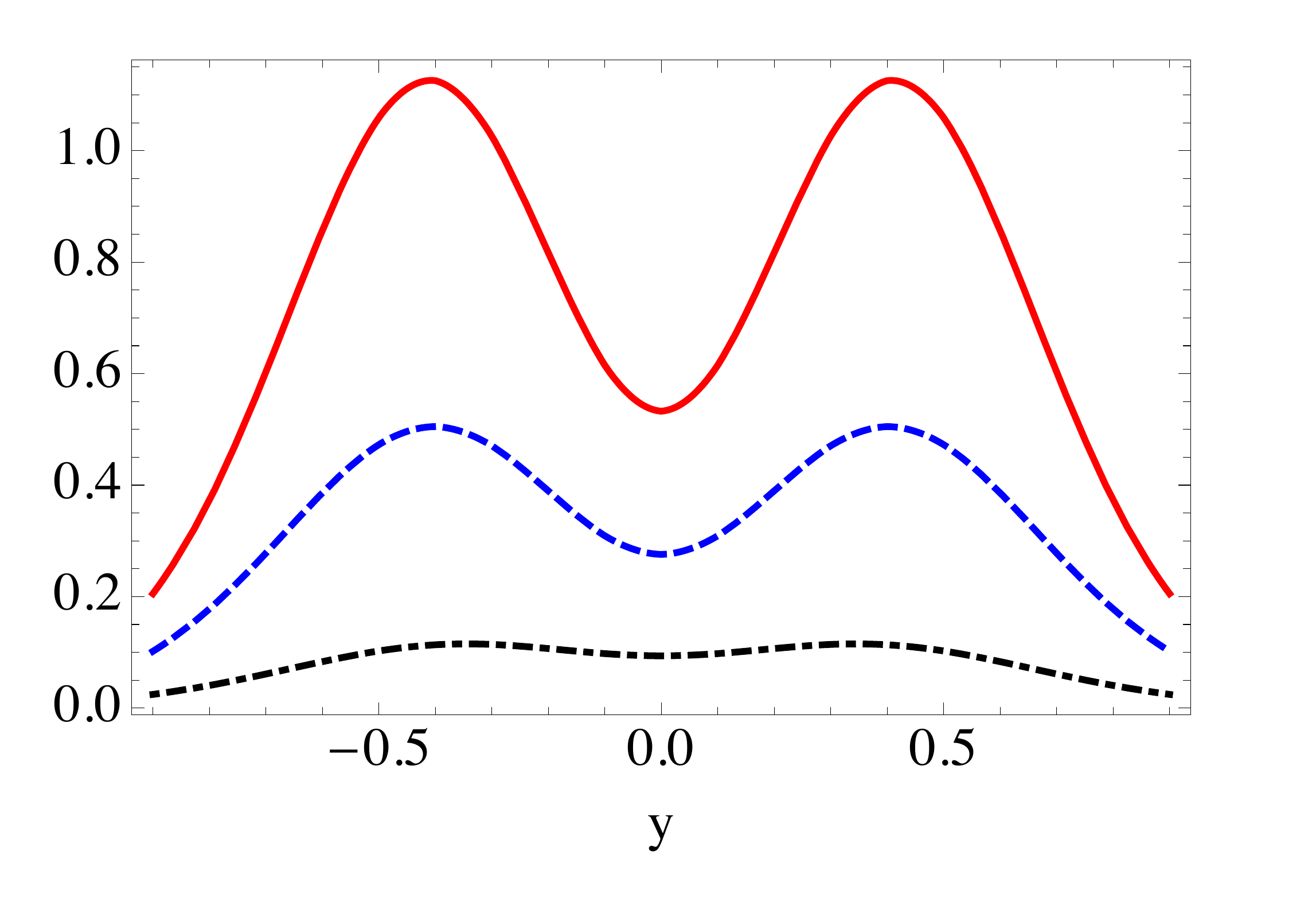}
 \caption{
 Single particle distribution  as a function of rapidity y, for azimuthal angle $\Phi$=0, 0.3, 0.6 (red,blue, black)
 counted from the cluster location. 
 }
 \label{fig_single} 
 \end{figure}

The integrals over r and $\eta$ are computed using Mathematica's numerical integration, while we approximated the integral over $\varphi$ by using a well known saddle point method $\phi=\phi_p$. The pseudorapidity integral was evaluated in the range $|\eta|<5$, while the integral over r was calculated from $r=0$ to $r=r_f$, where $r_f$ is the value of the radius at which the background $v_{\perp}$ on the freeze-out surface reaches its maximum. A sample of the results is shown in Fig.\ref{fig_single} for pions at $p_{\perp}=1.5$ Gev: here one can also find the characteristic double-peak shapes.

Here however comes the difficulty: in the theoretical calculation we may calculate all distribution knowing the location of the  original cluster. In particular,  in Fig.\ref{fig_single} the angle is counted from the cluster location.  In experiment cluster location in azimuth and rapidity $\phi_c,y_c$
are unknown, and thus we can only observe correlators {\em integrated over them}. Reconstructing from those the original single-body distribution is not a trivial task.

In principle, in order to solve the case, an experiment should be able to measure a sample of $n$-body correlation functions.
Returning to the rapidity case at hand, those can be written as
\be  {dN \over dy_1\ldots dy_n} = \int dy_c P(y_c) \prod_{i=1..n} f(y_i-y_c) \, ,
\ee 
where $P(y_c)$ is the probability to have a cluster at rapidity $y_c$, and $f(y_i-y_c)$ being spectrum modification
due to perturbation which we just calculated above. 

As rapidity distributions are usually rather rapidity-independent,  $P(y_c)\approx const$ can be approximated by a constant. If so,   the  $n$-body distribution depends on rapidityˆ differences. Furthermore, translational symmetry in rapidity results in conservation of the momentum associated with this coordinate. One obvious consequence is that the two particle correlation is thus a function of $y_1-y_2=\Delta y$. Furthermore, it is convenient to define Fourier transform
 \be \tilde f(k)=\int dy e^{iky} f(y) \, ,\ee
and rewrite the Fourier transform of the n-body spectrum in a form
\be  {d\tilde N \over dk_1\ldots dk_n} =2\pi \delta(\sum_{i=1..n}k_i) \prod_{i=1..n}  \tilde f(k) \, ,\ee
where the delta function stems from the integral over unknown cluster rapidity $y_c$.
A very special case is 2-body one, in which there remains only one momentum since
$k_2=-k_1=k$ and one can rewrite this expression as the ``power spectrum"
\be  {d\tilde N \over dk}\sim |\tilde f(k)|^2 \, ,
\ee
containing the square modulus of the harmonic amplitudes, but not the phase. 

 As a particular toy model, consider $f(y)$ of a double-peaked shape, as we found in certain kinematic window.
 Representing it by $f(y)=(\delta(y-a) + \delta(y+a))/2$ one finds $\tilde f(k)= cos(ka)$, and the power spectrum thus
 being \be |\tilde f(k)|^2=(e^{2ika}+ e^{-2ika}+2)/4 \, .\ee
 Making a Fourier transform of the power spectrum one finds the 2-particle correlation function:
 three terms in this expression giving three peaks, at $\Delta y=\pm 2a$ and at $\Delta y=0$, of twice larger amplitude.
 
 This issue and expressions are the close analogs of formulae which had been derived in the theory of correlators as a function of the  azimuthal angle. In particular, a three-peak structure of the kind had emerged from hydrodynamical calculation in our work \cite{Staig:2011wj}. Indeed, for central collisions one has the axial symmetry of the background flow, resulting in angular momentum zero for observable harmonics of the any-body correlators.

(As a side remark, we point out that  while  the experimental two-body correlator
does indeed have the predicted shape with three maxima, that does $not$ uniquely prove that the original spectrum has indeed the predicted shape. For example, various harmonics may have random phases, which are not observable in the power spectrum. This issue for flow harmonics remains unresolved and needs further study of few body correlators.)
 
Summarizing this part: given the single-particle perturbation function $f(y)$, all multi particle ones can be calculated, e.g. from the (approximate) relations above. We however cannot offer any straightforward inverse procedure,
deriving  $f(y)$ from measured correlators: comparing calculated predictions with the measurements seem to be the only way. Since there are many  multi body correlation functions, one should be able eventually get to convinced
that  $f(y)$ have certain  shape, such as e.g. the one coming from the  projected sound sphere. Certainly, quantifying the multibody correlations needs a lot of statistics: to characterize what can be done with the experimental sample on tape let us just mention that studies of the elliptic flow factorization already done include up to $n=2,4,6,8$ particles. 
 
Let us now return to joint 2-particle distributions, both in rapidity and angle    
\begin{align}
\frac{dN}{d\Delta\phi d\Delta\eta} &= \int \frac{dN}{d(\eta_1 - y)d(\phi_1 - \psi)}\frac{dN}{d(\eta_2-y)d(\phi_2-\psi)}d\psi dy \
\end{align}
Unfortunately, as one can see from Fig.\ref{tauc_taufo}, the time  
 for sound propagation under consideration is rather limited to about 2 fm,
 except in the improbable case of a cluster at very large $r>6\, fm$. Thus 
  the sound-induced peaks in rapidity are only shifted by about $\pm 1/2$ unit of rapidity, which in most kinematics is not enough to see the peaks in  the observable correlators.
  
\begin{figure}[!t]
\includegraphics[width=7cm]{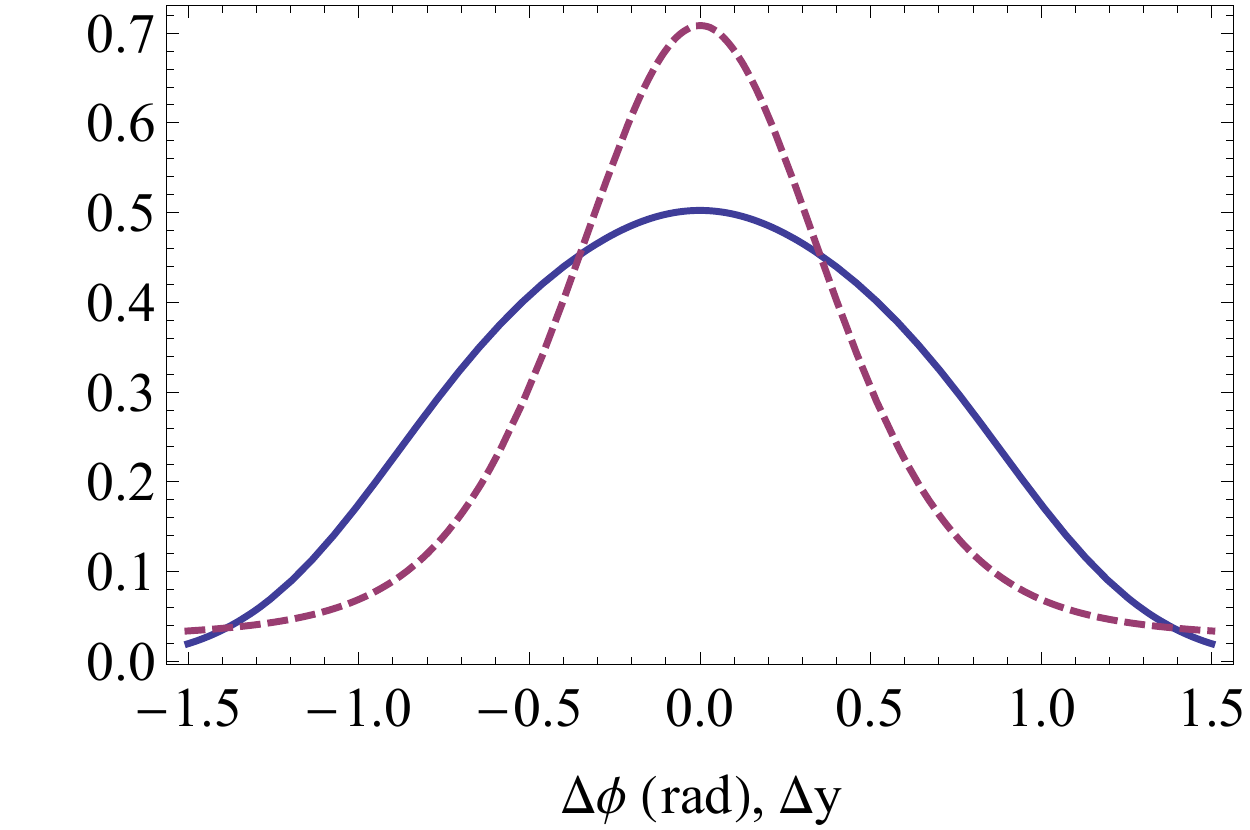}
\includegraphics[width=7 cm]{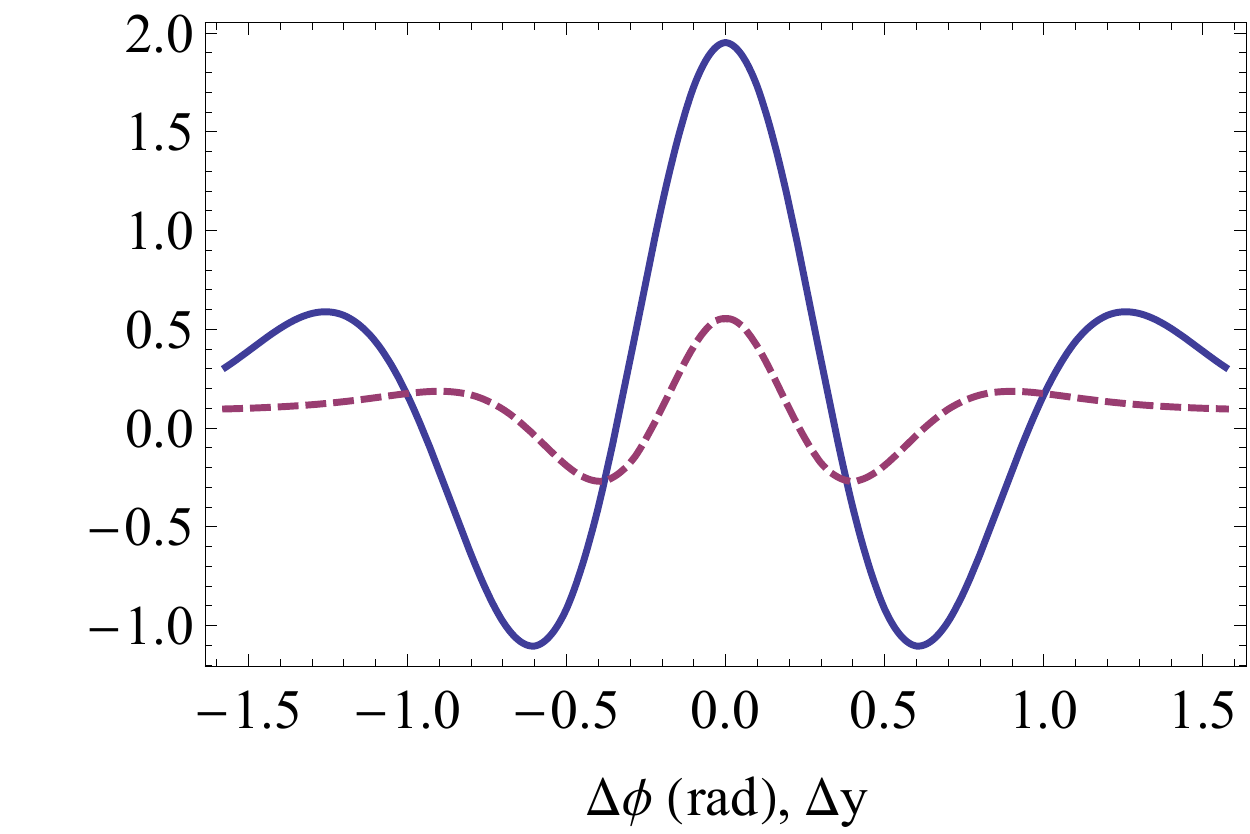}
 \caption{Two particle correlation functions for a cluster located initially at: (Top)$r=6$ fm, and (Bottom) $r=6.5$ fm. The solid (blue) curve corresponds to the correlation in $\Delta\eta$ with $|\Delta\Phi|<0.87$, and the dashed (magenta) curve corresponds to the correlation in $\Delta\Phi$ with $|\Delta\eta|<0.8$. In both cases the single particle distribution functions had been normalized such that their integral was 1.}
 \label{tpc} 
 \end{figure}

   After studying those, we come to the conclusion that
most interesting seem to be an asymmetric kinematics, in which the trigger is higher momentum  (or higher mass) and serve to locate the cluster location, while the associate particles should rather be a pion with smaller $p_\perp$, sensitive to the double-hump region of the fireball.
We calculate the two particle correlation with one particle with $p_\perp=1.5$ GeV and one with $p_T=2.5$ GeV. We integrate over $\Delta\phi$  in the range $|\Delta\phi|<0.87$, to obtain the particle correlations projected in $\Delta\eta$, and integrate over $|\Delta\eta|<0.8$, to generate the two particle correlation projected in $\Delta\Phi$.  We present the results in Fig. \ref{tpc} for two initial locations of the perturbation, at $r=6$ fm and $r=6.5$ fm. As one will see,  the correlations obtained are very different, which is explained by a different
time for the sound to propagate.  
The shape of the particle distributions and the two particle correlations shown in Fig. \ref{tpc} varies greatly, depending on the initial radial position of the perturbation. For perturbation at   located at at $r=6$ fm one fins all three peaks in the correlators merge into one structure,
while for that at   $r=6.5$ fm one can clearly see the three peaks.
This happens, again,  because for the different sound origination points, the evolution will be longer at some places and shorter at others.  Furthermore, to produce a noticeable effect the perturbation must be placed near the edge of the fireball---if it is located close to the center the sound circles will not reach the edge.

\subsection{Phenomenology}
\label{sec_pheno}

Correlations of secondaries in rapidity is a subject which, in the context of $pp$ collisions, goes back at least to 1970's experiments at CERN ISR. Already at that time it had been recognized that secondary pions are not produced individually, but from certain $clusters$ of $\langle N_{ch}\rangle \sim 3$ charged secondaries, or about 5 pions. Their mass and apparent isotropic decay distribution (deduced from shape and width of the correlation function itself) indicated that they are some hadronic resonances, with  the mass  $M\sim 2 \, GeV$. With the development of string fragmentation models -- such as Lund model and its descendants like Pythia -- these observations were naturally explained.

Heavy ion collisions at RHIC have not  focused so much on rapidity correlations. We have only PHOBOS collaboration data \cite{Alver:2008aa}, which used their large rapidity coverage due to the silicon detector. (PHOBOS had no particle ID or momentum measurements, so the pseudo rapidity has been used.)   These data display  rather strong modifications of the two-particle correlators in AA, relative to pp.   Analysis of those data using some version of a cluster model has been reported by G.S.F.Stephans in the talk \cite{Stephans}. Their discussion and some key plots have been reproduced in ref.\cite{Shuryak:2009cy}, so we will not duplicate it here and only summarize the main points. In Fig.13(a) of \cite{Shuryak:2009cy},  from Stephans, one can see that the charged multiplicity per cluster  in AuAu collisions  is significantly larger that what is seen  pp collisions, up to  $\langle N_{ch}\rangle \sim 6$ charged particles (or up to 10 including neutrals). Furthermore, as shown in Fig.13 (b), the produced clusters do not decay isotropically but are instead more extended in (pseudo)rapidity. The width of the cluster decay changes from about 0.8 in pp to about 1.4 at mid-central collisions, a quite substantial broadening. The first fact might bring to mind production of heavier  resonances, but the last feature  excludes this, as the decay of resonances can hardly be anisotropic.

\begin{figure}[!t]
 \center{ \hskip 0in\includegraphics[width=7cm]{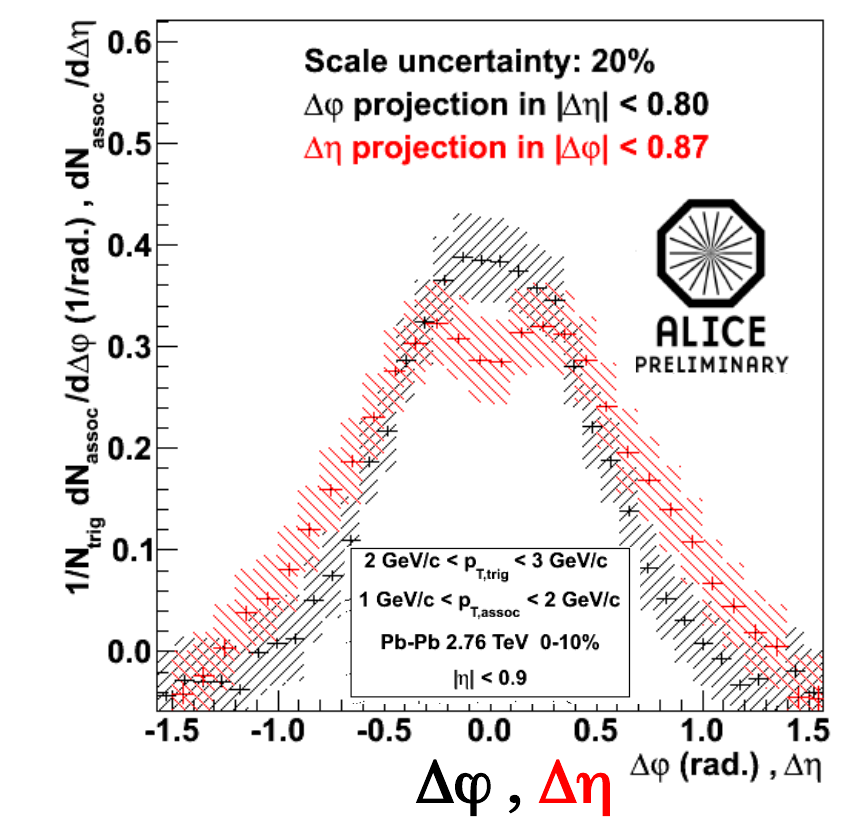}}
 \caption{ 
 Correlation functions of two charged hadrons in the kinematic range defined in the figure,
 as a function of pseudorapidity and azimuthal angle differences between the two particles \cite{Alice_corr}
 \label{fig_ALICE} }
 \end{figure}

The first LHC data on two-particle pseudorapidity correlations provided further puzzles.  As seen in Fig.~\ref{fig_ALICE} (from ALICE collaboration \cite{Alice_corr}), the  observed correlator seem to have the two-hump shape. (The evaluated kurtosis of this distribution is near -1, with rather high statistical significance away from zero: so a shape change cannot be a 
statistical fluctuation.). This particular ALICE 
plot is kinematically restricted to rather high $p_t$ of both trigger and associate particle, so one cannot compare it to PHOBOS data directly.  (One may wander if the two humps is not due to the well known Jacobian between the rapidity and psudorapidity. In Appendix we show it not to be the case, with the  Jacobian contributing an effect of one order of magnitude smaller than observed.)

Based on our arguments above, we propose  the following interpretation of these phenomena:\\ 
(i) the increased width in rapidity and the modified shape are caused by the sound waves emitted by the decaying clusters.\\
(ii) the larger cluster mass is due to larger QGP clusters in the hadronic matter produced in the  AA case, substituting the string fragmentation
process in the pp collisions\\   

Our calculations above had produced a variety of shapes of the two-particle correlators, from near-Gaussian to three-peak ones.
Some combination of those can perhaps generate the shape seen in experiment.

While we argued above that known examples of cluster collapse lead to efficient (nearly complete) transfer of its stored energy into the shocks/sound: but in practice the efficiency of this process is hard to evaluate. 
 If one assumes a two-component model of the particle source, in which certain number of secondaries, proportional to extra parameter $A$, originate from the QGP cluster itself, 
 while a number proportional to parameter $B$ come from the sound emitted from its collapse.
  The two-particle correlator,  projected on the rapidity $difference$ $\Delta y$, is then written  in a form  of three terms 
\be {dN_{corr} \over d \Delta y}= A^2  f_{cc}( \Delta y) + AB f_{cs}( \Delta y ) +B^2 f_{ss}( \Delta y )
\ee 
The first term stands for both secondaries coming from the cluster decay: as it is expected to decay isotropically the function $ f_{cc}( \Delta y)$ is the same near-Gaussian distribution as is well known for the two-body resonance decays. The second term has a trigger coming   from the cluster, fixing its rapidity, and the second from the sound: we thus expect the function
$f_{cs}( \Delta y )$ to have the double-hump shape we have calculated in the preceding section. 
The third term $\sim B^2$  is the convolution of the two single-particle ones just specified, averaged over the unobserved rapidity of the cluster, calculated in the preceding section. 
Unfortunately, if $A$ is non-zero, such a modification leads to reduction of the correlator width and even less opportunity
to get a double-hump shape. 

\subsection{Can late-time sounds be observed via higher angular harmonics?}
The calculated two-particles correlators shown 
in Fig. \ref{tpc} display certain structures not only in rapidity, but in the azimuthal angle as well, with the characteristic width
$\delta \phi \sim 1 \, rad$. Those would correspond to angular harmonics $m\sim 2\pi /\Delta \phi\sim  6$ and higher.
We had already mentioned in the Introduction that the {\em so far} observed multiple harmonics in azimuthal angle $\phi$ come from the initial
time perturbations. Their strength is peaked at $m=2,3$, but the existing data do extend at least till $m=6$. 

 Since the damping factor \cite{Staig:2010pn}
\be {v_m^2 \over \epsilon_m^2} \sim exp(- {4\over 3} {\eta \over Ts} {m^2 t  \over R^2} )\ee
exponentially  decreases with the time of propagation $t$, one may argue that at large enough
harmonic number  $m>m_{late}$ the late-time sounds become dominant over the early-time ones.
While the initial-state fluctuations should travel the time till freeze out $t_f$, the critical sound should only
propagate time $t_f-t_c$. The equation for $m_{late}$ then becomes
\be \epsilon_m^{initial} e^{-m_{late}^2 {2\over 3} {\eta \over Ts R^2} t_f } <  \epsilon_m^{c} e^{-m_{late}^2 {2\over 3} {\eta \over Ts R^2}( t_f-t_c) }
\ee
or 
\be   m_{late}\approx {3\over 2} {T R^2 \over t_c}    {s \over \eta} ln({\epsilon_m^{c} \over  \epsilon_m^{initial}}) \ee

The situation is complicated further by the fact that higher order harmonics can be generated also non-linearly,
as a superposition of several lower harmonics. (E.g. m=6 can be generated as 2+2+2 or 3+3.)
In this case the amplitude of the signal is reduced, but also the damping effect is
less severe. 
At the moment it is hard to see, if the progress in experimental statistic/accuracy may get sufficient
to  find evidences for the ``let time sounds" in the angular harmonics as well.

\section{Summary and discussion}
In this paper we  \\
(i) have assumed that during passing of the $T\approx T_c$ region of the QCD phase transition some inhomogeneous intermediate state of matter is reached,
resulting in formation of the ``QGP clusters" ;\\
(ii) had shown that they likely to undergo the Rayleigh collapse, as the hadronic phase pressure becomes higher than that of the QGP\\
(iii) This collapse converts significant fraction of the cluster's energy into an outgoing shock/sound pulse (the mini-bang)\\
 (iv) which propagates, and by the time of the final freeze-out (with about  $\delta\tau\sim 3 fm$ to go) generate sound spheres
 of the size $c\delta\tau \sim 1.2 fm$.
 
We further propose that increases in clustering and especially of their rapidity width and modified shape
observed by PHOBOS at RHIC and ALICE at LHC can be the manifestation of these
late-time sound spheres.

 Needless to say, a lot of studies need to be done before these suggestions can be verified.
In particular, our arguments rely on hydrodynamics: but the famous ``perfect liquid"
properties  of the matter are known for QGP, not so much for 
the late-stages hadronic matter.  

Another needed disclaimer is needed for hydrodynamical expressions we use.
Gubser flow is a very attractive analytic tool, saving us months of numerical studies:
yet it is not realistic enough and can at best be used for the QGP stage of the collision, not the
late hadronization stages we used it for in this paper. So, 
 the particular results we obtained cannot be trusted  beyond 
a qualitative level.


\section{Appendix A: The Jacobian dip } 
There is the so called ``Jacobian dip" in the pseudorapidity
 $\eta=(1/2) \ln((p+P_l)/(p-P_l))$ distribution as opposed to true
rapidity $y=(1/2) \ln((E+P_l)/(E-P_l)) $: indeed 
\be {dy \over d\eta} = {1 \over \sqrt{1+m^2/(p_t \cosh(\eta)^2}} \ee 
 but neither the magnitude nor the width of the observed dip can be explained by it.
 
 {\bf Acknowledgments.} 
 supported in parts by the US-DOE grant
DE-FG-88ER40388.

\end{document}